\documentclass[english,journal,final]{IEEEtran}
\usepackage[T1]{fontenc}
\usepackage[latin9]{inputenc}
\usepackage{color}
\usepackage{verbatim}
\usepackage{graphicx}

\makeatletter
\author{Junmo Sung, Mostafa Sayed, Mahmoud~Elgenedy, Brian L. Evans, Naofal Al-Dhahir, Il Han Kim, and Khurram Waheed
\thanks{J. Sung and B. L. Evans are with the Department of Electrical and Computer Engineering at The University of Texas at Austin, Austin, TX 78712 (e-mail: \{junmo.sung, bevans\}@utexas.edu).}
\thanks{M. Sayed is with Samsung Semiconductor, Inc., San Jose, CA 95134 (e-mail: Ms.ibrahim@samsung.com).}
\thanks{M. Elgenedy and N. Al-Dhahir are with the Department of Electrical Engineering, University of Texas at Dallas, Richardson, TX 75080 (e-mail: \{mahmoud.elgenedy, aldhahir\}@utdallas.edu).}
\thanks{I. H. Kim is with Texas Instruments Inc., Dallas, TX 75243 (e-mail: il-han-kim@ti.com).}
\thanks{K. Waheed is with NXP Semiconductors, Austin, TX 78735 (e-mail: khurram.waheed@nxp.com).}
\thanks{This paper is based in part on the PhD thesis of Dr. Mostafa Sayed Ibrahim titled "Hybrid Powerline/Wireless Diversity for Smart Grid Communications.", The University of Texas at Dallas, 2018.}
\thanks{Fig. 1 was created by Dr. Jing Lin (UT Austin) and used with permission.}}

\usepackage{cite}
\usepackage{balance}
\usepackage{url}
\usepackage{csquotes}
\usepackage{fixltx2e}

\usepackage{babel}

\@ifundefined{showcaptionsetup}{}{%
 \PassOptionsToPackage{caption=false}{subfig}}
\usepackage{subfig}
\makeatother

\usepackage{babel}
\begin{document}

\title{Hybrid Powerline/Wireless Diversity for Smart Grid Communications:
Design Challenges and Real-time Implementation}

\maketitle
The demand for energy is growing at an unprecedented pace that is
much higher than the energy generation capacity growth rate using
both conventional and green technologies.In particular, the electric
power sector is consistently rated among the most dynamic growth markets
over all other energy markets. Distributed (decentralized) energy
generation based on renewable energy sources is an efficient and reliable
solution to serve such huge energy demand growth \cite{IEO2016report}.
However, to manage dynamic and complex distributed systems, a massive
amount of data has to be measured, collected, exchanged and processed
in real time. Smart grids manage an intelligent energy delivery network
enabled two-way communications between data concentrators operated
by utility companies and smart meters installed at the end users.
In particular, dynamic power-grid loading and peak load management
are the two main driving forces for bidirectional communications over
the grid. Narrowband power line communications (NB-PLC) and wireless
communications in the unlicensed frequency band (sub-1 GHz or 2.4
GHz) are the two main communications systems adopted to support the
growing smart grid applications. Moreover, since NB-PLC and unlicensed
wireless links experience channel and interference with markedly different
statistics, transmitting the same information signal concurrently
over both links significantly enhances the smart grid communications
reliability. In this article, we compare various diversity combining
schemes for simultaneous power line and wireless transmissions. Furthermore,
we developed a real-time testbed for the hybrid PLC/wireless system
to demonstrate the performance enhancement achieved by PLC/wireless
diversity combining over a single link performance.

\section{Introduction}

\begin{figure}
\begin{centering}
\includegraphics[angle=-90,scale=0.32]{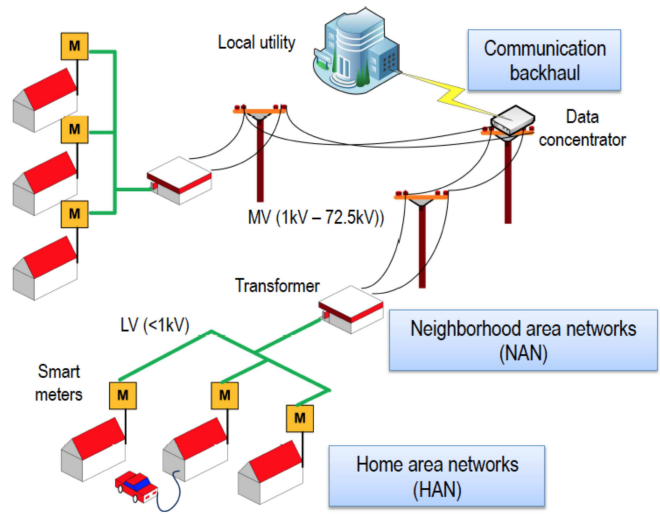} 
\par\end{centering}
\caption{\label{fig:V2G-System-Architecture-1}Smart grid communication networks.}
\end{figure}

A smart grid couples a two-way communication network to the traditional
power grid to enable adaptive energy management. For smart metering
applications, a smart grid consists of three primary communication
networks, namely, home area network (HAN), neighborhood area network
(NAN) and backhaul communications network as depicted in Fig. \ref{fig:V2G-System-Architecture-1}.

\begin{comment}
\begin{itemize}
\item Home area networks (HAN) that connect smart appliances at homes and
business locations as well as the sensors deployed over indoor powerlines
to smart meters; 
\item Neighborhood area networks (NAN) that interconnect smart meters scattered
within the grid and data concentrators that are deployed by local/regional
utility companies over medium-voltage (MV) powerlines (in the US)
or low-voltage (LV) powerlines (in Europe); and
\item A backhaul communications network that connects data concentrators
to local/regional utilities. 
\end{itemize}
\end{comment}

Power line and wireless communication technologies are two important
candidates that support smart grid communications \cite{galli2011grid}.
Next, we provide an overview of power line and wireless communication
fundamentals.

\subsection{Powerline Communications System}

PLC is an appealing solution for communications in HANs and NANs considering
its low deployment cost over existing infrastructure. However, PLC
must overcome several challenges to provide a reliable communication
link. For example, the PLC channel is highly frequency selective and
experiences instantaneous changes due to dynamic switching and branching
in power lines \cite{nassar2012local}. In addition, a typical PLC
system suffers from high interference and impulsive noise that dominate
the background noise power and can result in severe performance degradation.
In PLC, interference and impulsive noise are mainly generated by electrical
devices connected to the power line grid. An additional source of
interference is caused by external signals coupled to the power lines
through conduction or radiation \cite{galli2011grid}. Based on the
operating frequency bands, there are three categories of PLC systems
\cite{nassar2012local,galli2011grid}:

\begin{comment}
In particular, PLC can work where radio propagation is weak or completely
blocked. For example, a data concentrator in the basement of a building
can barely send or receive wireless signals, while PLC can traverse
the power lines to reach smart meters in apartment units upstairs.
\end{comment}

\begin{itemize}
\item Ultra-Narrowband power line communications (UNB-PLC) systems that
operate in the frequency band of $0.3-3$ kHz to support around $100$
bps data rate for distances more than $150$ km. UNB-PLC used by utilities
for supervisory control and data acquisition of the power generation
units.%
\begin{comment}
\begin{itemize}
\item An example of such systems is the Two-Way Automatic Communications
System (TWACS) that has been deployed by hundreds of utilities over
the last two decades for automatic meter reading, outage detection
and voltage monitoring \cite{nassar2012local,galli2011grid}. 
\end{itemize}
\end{comment}
\item Narrowband power line communications (NB-PLC) systems in the $3-500$
kHz frequency band for data rates up to several hundred kbps using
Orthogonal Frequency Division Multiplexing (OFDM). Recently, NB-PLC
gained significant interest to support NANs. Industry developed standards
for NB-PLC include G3, PRIME, IEEE $1901.2$ and ITU-T G.hnem. %
\begin{comment}
\begin{itemize}
\item The former, a.k.a. low data rate NB-PLC, has been used for commercial
building automation (e.g. automatic lighting and air conditioning),
and has been standardized in BacNet and LonTalk \cite{nassar2012local,galli2011grid}.
In fact, it is currently the most adopted ($60$\% market share) communication
technology in smart meters. , including the European CENELEC (Committee
for Electrotechnical Standardization) bands ($3-148.5$ kHz), the
US FCC (Federal Communications Commission) band ($10-490$ kHz), the
Japanese ARIB (Association of Radio Industries and Businesses) band
($10-450$ kHz) and the Chinese band ($3-500$ kHz).
\end{itemize}
\end{comment}
\item Broadband power line communications (BB-PLC) systems operate in the
$1.8-250$ MHz frequency band providing several hundred Mbps data
rates to support HANs. Standards for BB-PLC such as TIA-$1113$ (HomePlug
$1.0$), ITU-T G.hn and IEEE $1901$ specifications.
\end{itemize}
In this article, we focus mainly on the NB-PLC for smart grid communications.
In NB-PLC, apart from the interference caused by instantaneous switching
elements, the dominant interference is a cyclostationary impulsive
noise synchronous to half of the AC cycle.

\begin{comment}
Typical sources of the noise and interference include non-linear power
electronic devices such as inverters, DC-DC converters, and long-wave
broadcast stations whose energy is coupled to the power lines in the
$3-500$ kHz band.
\end{comment}

\subsection{\textcolor{black}{Wireless Communication System}}

Initially, cellular communications was the main wireless communication
system used for smart meters communications application. The main
drawback of cellular technologies is the high running cost of leasing
networks/services from the carriers. Later, wireless mesh networks
gained much attention as a low-power/low-priced solution for the application
of smart metering communications. International standards for mesh
networks include IEEE 802.15.4 O-QPSK (used by Zigbee, Z-wave, Thread,
etc.), IEEE $802.11$ah and IEEE 802.15.4g. Specifically, to connect
smart meters to data concentrators, ZigBee technology was used to
deliver $20-250$ kbps in the frequency bands around $868$ MHz, $915$
MHz or $2.4$ GHz over a distance of $10-200$ m. Moreover, wireless
smart utility networks (Wi-SUN) in the frequency band of $450$ MHz-$2.4$
GHz, based on the IEEE $802.15.4$g standard, provide different data
rates to support low-power indoor communications between smart meters
and smart appliances. Furthermore, the IEEE $802.11$ah standard supports
a few hundred kbps data rates over $200$ meters in the sub-$1$ GHz
unlicensed frequency bands targeting smart metering applications.

The main challenge for wireless communications over unlicensed frequency
bands is the existence of strong interference caused by uncoordinated
transmissions. Specifically, neighboring devices based on different
standards running in the same frequency band cause interference to
each other. Such interference is impulsive and can be described statistically
using different models including the Middleton Class-A, Gaussian mixture
(GM), and symmetric alpha stable models \cite{nassar2011low}. Henceforth,
the term \textquotedbl{}noise\textquotedbl{} is used to refer to both
thermal noise and interference.

\section{Hybrid PLC-Wireless System Model\label{sec:System-Model}}

As discussed earlier, the basic problem in smart grid communications
is the existence of high interference which severely degrades the
communication reliability. Such strong interference motivates utilizing
both PLC and wireless communication systems for concurrent transmission
of the same data signal. At the receiver side, the PLC and wireless
signals are combined to realize transmission diversity gains. An important
advantage of the hybrid PLC/wireless system diversity is the statistical
independence of the channel and interference on both links, which
is referred to herein as asymmetric diversity. This is in contrast
to the conventional symmetric PLC/PLC or wireless/wireless diversity
systems where the channel and interference are generally correlated
across the diversity branches and have the same statistics. In particular,
the channel/interference high correlation in PLC/PLC system results
from the close proximity between the power lines. Similarly, in wireless/wireless
diversity, the channel/interference correlation can be also high since
the antenna separations are likely to be smaller than half the wavelength
in the $900$ MHz frequency band. Therefore, for hybrid PLC/wireless
systems, new diversity combining techniques are needed to exploit
the asymmetric channel and interference characteristics of the PLC
and wireless links.

\begin{figure*}[t]
\begin{centering}
\includegraphics[scale=0.42]{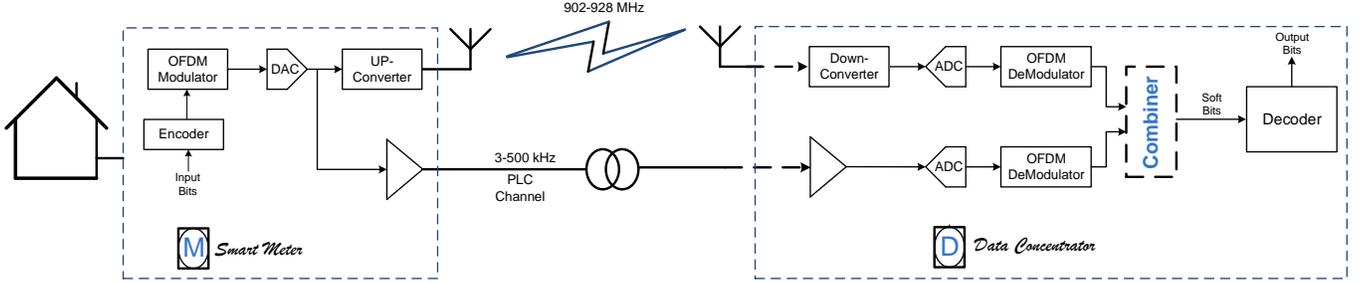} 
\par\end{centering}
\caption{\label{fig:System-Block-Diagram}Hybrid PLC/Wireless System.}
\end{figure*}

The hybrid PLC/wireless system is shown in Figure \ref{fig:System-Block-Diagram}
where OFDM is adopted for NB-PLC and unlicensed wireless (sub-$1$
GHz) standards. At the transmitter, the same data is simultaneously
transmitted over PLC and wireless links. At the receiver, signals
received on both links are combined based on log-likelihood ratios
(LLRs). In particular, the LLRs (soft bits) are calculated separately
for each link and then combined (added) using proper weights. To estimate
the transmitted information signal, the combined LLRs are fed to the
channel decoder. Note that LLR combining is done at the bit-level
which allows the PLC and wireless links to use different parameter
settings such as the size of the fast Fourier transform, constellation
size and cyclic prefix length, assuming the same bit rate for both
links.

\subsection{NB-PLC Noise and Channel Models\label{subsec:Noise-and-Channel}}

The impulsive noise in the NB-PLC link is modeled as a cyclostationary
random process with a time period equal to half of the AC cycle \cite{nassar2012cyclostationary,elgenedy2016Cyclostationary}.
The noise model of \cite{nassar2012cyclostationary} splits the cyclostationary
noise period into multiple temporal regions where the noise in each
region is assumed a stationary random process. Moreover, every temporal
region is generated using a linear time-invariant filter estimated
from experimental field measurements. In \cite{elgenedy2016Cyclostationary},
frequency-shift filters are used to shape the spectrum of a stationary
white noise signal into a cyclic spectrum constructed based on experimental
field measurements.

The NB-PLC channel response depends mainly on the power line network
topology including the different connected electrical devices. Transmission
line (TL) theory is used to model the NB-PLC channel {[}8{]}. A detailed
analysis of using TL modeling to characterize the channel responses
of measured field data is given in the IEEE P1901.2 standard.

\begin{comment}
PLC channel modeling based on field measurements is usually adopted
in order to make sure that the predicted performance is close to the
practical scenario .
\end{comment}

\subsection{Unlicensed Wireless Noise and Channel Models}

There is a lack of research studies on noise modeling in the unlicensed
wireless sub-1 GHz frequency band ($902-928$ MHz). However, transmission
in the sub-$1$ GHz frequency band is analogous to the $2.4$ GHz
frequency band since both bands are unlicensed with similar operating
communication standards, e.g., IEEE 802.11 and IEEE 802.15.4 wireless
standards. Specifically, Bluetooth, Wi-Fi and ZigBee operate in the
$2.4$ GHz frequency bands while Sub-$1$ GHz frequency bands occupants
include Wi-SUN, IEEE $802.11$ah, EnOcean, and ONE-NET standards.
Thus, similar to \cite{nassar2013graphical} which was introduced
mainly for the $2.4$ GHz frequency band, impulsive noise can be modeled
as a GM random process.

For the wireless link, to model non-line-of-sight propagation, we
used the Rayleigh fading channel model.

\section{PLC/Wireless Diversity Combining\textcolor{blue}{{} }for Coherent
Modulation\textcolor{blue}{\label{subsec:TD-Noise-Filtering}}}

\subsection{Average Signal to Noise Ratio Combining (ASC)}

As discussed earlier, the statistics of the impulsive noise in the
NB-PLC and wireless systems are independent. In addition, the NB-PLC
impulsive noise is not a stationary process. Therefore, deriving the
optimal maximal ratio combining (MRC) rule is quite challenging. Moreover,
calculating the sufficient statistic of the wireless signal given
the GM noise is very complicated \cite{haring2002performance}. Assuming
white Gaussian noise for both wireless and NB-PLC systems, a sub-optimal
MRC rule can be developed. In particular, the log-likelihood functions
of the received data symbols, per OFDM subchannel on each link, are
weighted by their corresponding average signal to noise ratio (SNR)
(which is the ratio of the squared channel gain per OFDM subchannel
over the average noise power) and then added to produce the combined
LL function.

\subsection{Instantaneous SNR Combining (ISC)}

The impulsive noise power levels on both the wireless and PLC links
change rapidly over the time and frequency domains. Consequently,
the average noise power metric considered in the ASC scheme is unreliable
and is a highly sub-optimal solution for this diversity combining
problem. Therefore, to capture the noise's instantaneous power changes,
the noise instantaneous powers per received data symbol are used in
the denominator of the combining weights \cite{sayed2014narrowband}.
\textcolor{black}{Key papers on estimating impulsive noise instantaneous
power for NB-PLC include \cite{lin2013impulsive,elgenedy_mitigation}.}

\subsection{Power Spectral Density Combining (PSDC)}

\textcolor{black}{Although the ISC scheme significantly outperforms
the ASC scheme, the computational complexity of estimating the noise
instantaneous power is very high}. Alternatively, the noise power
spectral density (PSD), or equivalently the average noise power per
OFDM subchannel, can be used for combining. Estimating the noise PSD
is much easier than estimating the instantaneous noise power as shown
in \cite{sayed2015ffficient}. \textcolor{black}{In particular, since
the NB-PLC cyclostationary noise can be modeled using multiple stationary
noise regions (i.e. the noise PSD is fixed per region) \cite{nassar2012cyclostationary},
the PSD of each temporal region can be estimated independently from
the received OFDM blocks if the noise region boundaries are known
(or previously estimated). }

\textcolor{black}{Techniques for estimating the PLC noise PSD are
presented in }\textit{\textcolor{black}{\emph{\cite{sayed2017diversity}}}}\textcolor{black}{{}
and \cite{lin2015time}. For example, in }\textit{\textcolor{black}{\emph{\cite{sayed2017diversity},
}}}\textcolor{black}{the noise PSD is estimated by first calculating
the average power of the received signal (per OFDM subchannel) and
then subtracting the average channel power from it. Moreover, to achieve
reliable estimates for the noise PSD, the time averaging duration
should be sufficiently long}\textit{\textcolor{black}{\emph{. It is
important to note that the NB-PLC channel is deterministic and is
either constant for all OFDM symbols or periodic with a period equal
to half the AC cycle. Thus, averaging the channel response over a
complete AC cycle is satisfactory to estimate the average channel
power per OFDM subchannel.}}}

\textcolor{black}{For the wireless link, the noise is already stationary
with a GM distribution. Therefore, all OFDM symbols can be used to
calculate the noise PSD. However, }\textit{\textcolor{black}{\emph{it
is shown in \cite{sayed2017diversity} that the wireless link noise
PSD is constant for all OFDM subchannels and is equal to the noise
variance. Thus, there is no need for noise PSD estimation for the
wireless link.}}}

\subsection{Joint Transmit-Receive Selection Diversity (TRSD)}

The PLC and wireless links are totally decoupled, i.e., the multi-input
multi-output (MIMO) channel matrix for the hybrid system (per OFDM
subchannel) is a $2\times2$ diagonal matrix. Therefore, the left
and right singular vectors of the channel matrix will be the columns
of the $2\times2$ identity matrix while the singular values will
be the absolute values of the channel coefficients per OFDM subchannel.
Therefore, to maximize the received SNR, the total transmit power
should be allocated to the medium with the higher channel-to-noise
ratio (CNR). Hence, transmit media selection is the optimal precoder
for the hybrid PLC/wireless transmission, unlike conventional PLC/PLC
or wireless/wireless systems which require complicated singular value
decomposition (SVD) precoding to diagonalize the cross-coupled channel
matrix. 

\begin{figure*}
\begin{centering}
\includegraphics[scale=0.53]{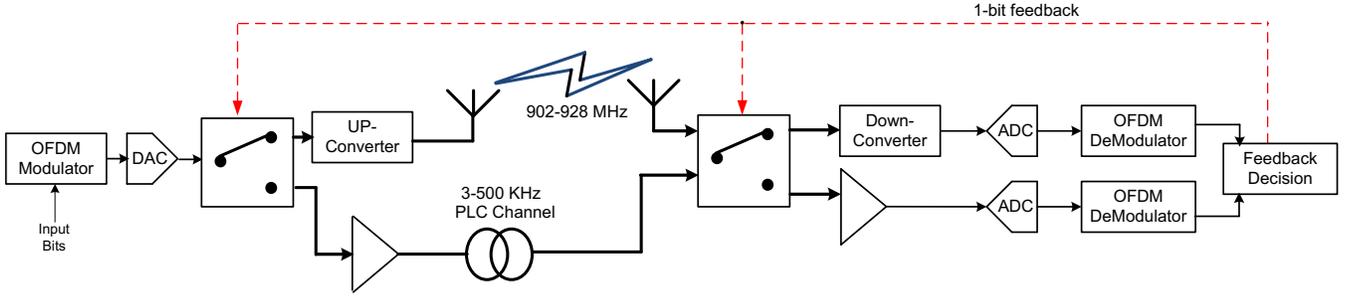} 
\par\end{centering}
\centering{}\caption{\label{fig:Block-diagram.-1}A Block diagram for the TRSD scheme.}
\end{figure*}

Figure \ref{fig:Block-diagram.-1} shows the TRSD block diagram where
a feedback link from the receiver to the transmitter is required for
the transmit selection operation. In contrast to the conventional
PLC/PLC or wireless/wireless transmit precoding, the TRSD scheme does
not require full knowledge of the instantaneous channel state information
(CSI) at the transmitter. However, only a single bit is required per
OFDM sub-channel to advise the transmitter on which link has the higher
CNR. Moreover, since the PLC link channel is deterministic, the feedback
rate is determined only by the wireless channel coherence time which
is typically large for smart grid applications with no mobility.

\section{PLC/Wireless Diversity Combining For Differential Modulation}

\subsection{Combining For Differentially-Modulated Wireless and NB-PLC Links}

Differential modulation is appealing to smart grid communications
because of its low design complexity. In particular, differential
modulation does not require channel estimation for detection, which
reduces the receiver design complexity significantly compared to the
coherent modulation receivers. In addition, differential modulation
is a mandatory scheme in the IEEE 1901.2 NB-PLC standard. In OFDM
systems, two types of differential modulation can be implemented,
namely, time-domain differential modulation (TDDM) and frequency-domain
differential modulation (FDDM). In TDDM, the data is transmitted in
the phase difference between two OFDM subchannels (at the same subchannel
index) of two successive OFDM symbols. However, in the FDDM, the data
is transmitted in the phase difference between two successive OFDM
subchannels in the same OFDM symbol. The choice of using TDDM or FDDM
depends on the channel characteristics, i.e., coherence time and coherence
bandwidth, respectively.

As an alternative to the SNR weighting scheme used in the coherent
modulation combining, the technique in \cite{sayed2016differential}
proposed to use the ratio of per subchannel absolute received signal
to the noise PSD as weighting factor for differential modulation combining.
This technique is referred to in \cite{sayed2016differential} as
differential signal strength combining (DSSC). For the differential
modulation case, a practical noise PSD estimation technique, named
``offline approach'', is discussed in \cite{lin2015time}. The DSSC
technique for differential modulation is shown in \cite{sayed2016differential}
to outperform the conventional ASC and Equal-Gain Combining (EGC)
techniques while the latter two techniques show similar performance.

\subsection{Combining For Coherently-Modulated Wireless Link and Differentially-Modulated
NB-PLC Link}

Wireless standards used for Smart Grids communications such as IEEE
$802.11$ah and IEEE $802.15.4$g do not include any differential
modulation schemes. Hence, a practical need arises to develop diversity
combining techniques to combine coherently-modulated signals from
the wireless link and differentially-modulated signals from the NB-PLC
link. In such scenario, the ratio of per subchannel absolute received
signal to the noise PSD can be used as a combining weight for the
PLC link with differential modulation. On the other hand, for the
coherent modulation used in the wireless link, the ratio between the
channel gain per subchannel over the noise variance per OFDM block
can be used as a weight in the combining metric.

\section{Hybrid PLC-Wireless Testbed Setup}

\begin{figure}
\subfloat[\label{fig:The-average-BER-1-2}BER performance for the BPSK vs $E_{b}/N_{o}$
of the PLC link at $E_{b}/N_{o}=3$ dB for the wireless link.]{\begin{centering}
\includegraphics[scale=0.47]{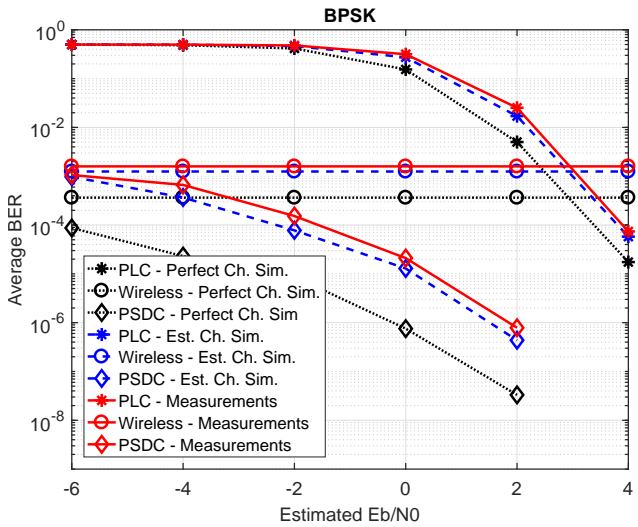} 
\par\end{centering}
}

\subfloat[\label{fig:The-average-BER-1-3}BER performance for the DBPSK vs $E_{b}/N_{o}$
of the PLC link at $E_{b}/N_{o}=3$ dB for the wireless link.]{\begin{centering}
\includegraphics[scale=0.47]{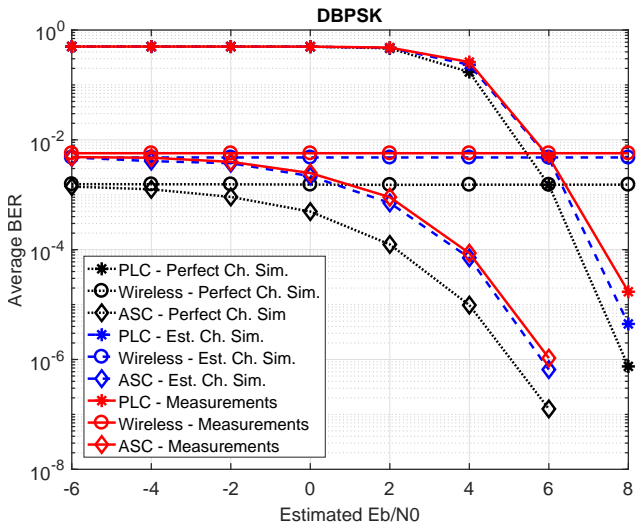} 
\par\end{centering}
}

\caption{\label{fig:The-average-BER-1-2-1-1-1} BER performance of the different
combining schemes, MATLAB simulation compared to the measured testbed
results.}
\end{figure}

To evaluate different combining techniques in practical conditions,
we developed a real-time testbed for the hybrid PLC/wireless system.
In addition, for the different diversity combining techniques, we
compare the testbed measured performance with MATLAB simulation results.In
Figure \ref{fig:The-average-BER-1-2-1-1-1}, we compare three types
of bit error rate (BER) curves : 1) MATLAB simulation assuming perfect
channel knowledge at the receiver side, 2) MATLAB simulation with
practical channel estimation, and 3) testbed measurements with practical
channel estimation. For each type, we show three BER curves: PLC system,
wireless system and PLC/wireless diversity combining. As examples,
for the PLC/wireless system performance curves, we use the PSDC technique
with coherent modulation while using the ASC technique with differential
modulation. As shown in Figures \ref{fig:The-average-BER-1-2} and
\ref{fig:The-average-BER-1-3}, for both the coherent and differential
BPSK cases, the testbed measured BER curves are very close to MATLAB
simulation BER curves (both with practical channel estimation). Moreover,
for the coherent BPSK modulation BER results shown in Figure \ref{fig:The-average-BER-1-2},
a gain of 3 dB in the $E_{b}/N_{o}$ at BER of $10^{-4}$ is achieved
by the hybrid PLC/wirless system using PSDC diversity combining over
the PLC system (assuming a constant $E_{b}/N_{o}=3$ dB for the wireless
link). For the differential BPSK, as shown in Figure \ref{fig:The-average-BER-1-3},
a similar $E_{b}/N_{o}$ gain over the PLC system is also achieved
by the hybrid PLC/wireless system using ASC diversity combining. Next,
we describe the developed testbed platform and design challenges in
more details. 

\begin{figure*}
\begin{centering}
\includegraphics[scale=0.45]{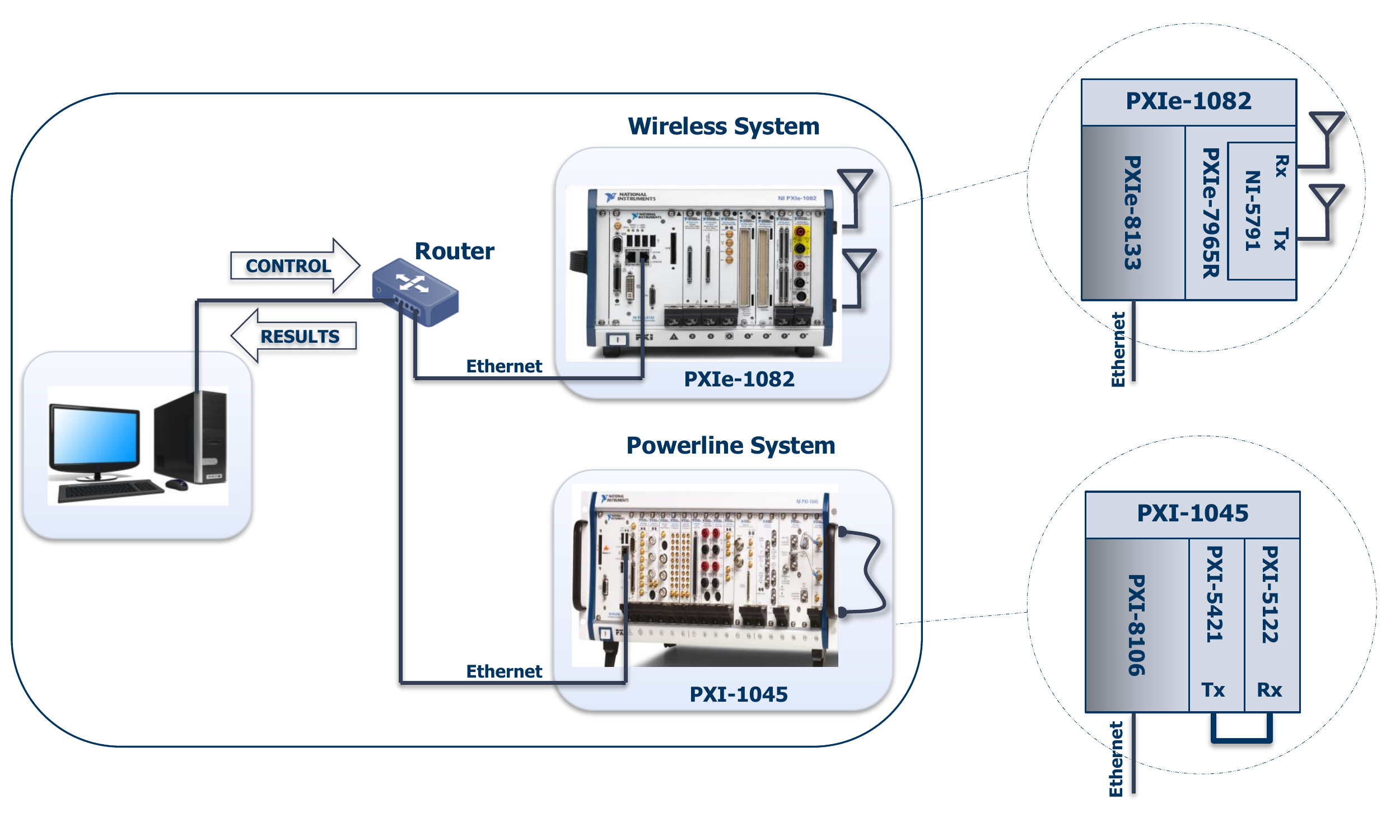} 
\par\end{centering}
\caption{\label{fig:The-average-BER-1}A block diagram for the testbed hardware
architecture.}
\end{figure*}

\subsection{Hardware Architecture}

As shown in Figure \ref{fig:The-average-BER-1}, the testbed is developed
mainly using National Instruments (NI) modules. In particular, the
PLC and wireless systems are implemented on two different chassis
and placed in different locations since each system has its own physical
channel environment. As shown in Figure \ref{fig:The-average-BER-1},
the PLC system is implemented on a PXI-$1045$ chassis that has different
slots to accommodate different modules such as a PXI-$5421$ signal
generator, a PXI-$5122$ digitizer and a PXI-$8106$ controller. Similarly,
the wireless system is implemented on a PXIe-$1082$ chassis which
consists of an NI-$5791$ RF adapter module, a PXIe-$7965$R FPGA
module and a PXIe-$8133$ controller. In addition, the RF adapter
module has both transmit and receive ports. Hence, a unidirectional
wireless link can be implemented using a single adapter module.

Finally, both controllers are connected to a personal computer (PC)
device through an Ethernet router so that the PC acts as a controller
and command center.

\begin{figure*}
\begin{centering}
\includegraphics[scale=0.52]{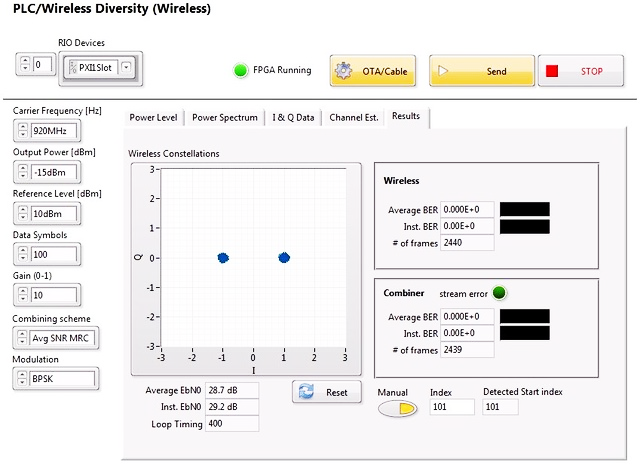} \includegraphics[scale=0.52]{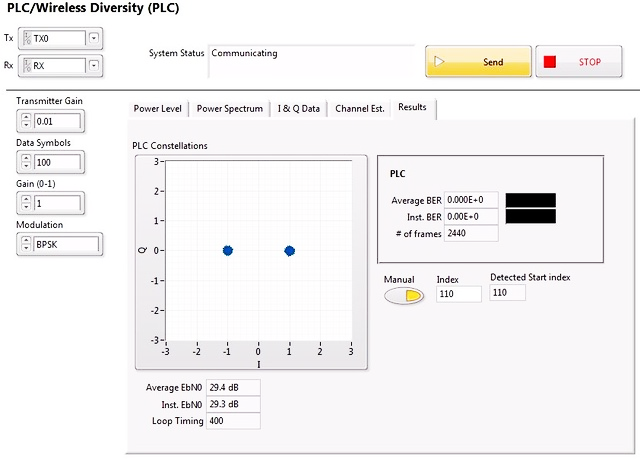}
\par\end{centering}
\caption{\label{fig:The-average-BER-1-1}The wireless and PLC systems software
front panels.}
\end{figure*}

\subsection{Software Architecture}

In the testbed, a real-time (RT) operating system (OS) runs on both
controllers while each controller has its own main program/front panel
to control/display system operations, as shown in Figure \ref{fig:The-average-BER-1-1}.
The PLC system controller runs one thread to perform frame generation,
bit decoding, and control signals transmission/reception with the
hardware. To test the implemented transceiver with or without channel
effects, the output signal from the transmitter can optionally bypass
the hardware and be directly forwarded to the receiver. Similarly,
for the wireless system, the other controller performs the same operations
with different hardware. In addition, the software threads are configured
such that the whole process is iterated at the same frame rate. Moreover,
in the wireless system controller, there is an additional thread that
performs the diversity combining operation. In particular, LLR outputs
from both the PLC and wireless threads are sent to this thread over
first-in first-out (FIFO) queues. For verification purposes, the additional
thread performs a comparison between the two bit streams to make sure
that the combining process is executed over the correct (synchronized)
frames.

\subsection{Packet Structure}

To apply the same packet detection schemes for both PLC and wireless
systems, the IEEE $1901.2$ standard's preamble format is used for
both PLC and wireless transmissions. Specifically, the preamble of
the IEEE $1901.2$ standard consists of $8$ or $12$ identical symbols
of type SYNCP, a full symbol of type SYNCM, and the first half of
a SYNCM symbol. SYNCP and SYNCM symbols are transmitted before the
frame control header (FCH) symbols where each symbol consists of $256$
samples. In addition, the SYNCP and SYNCM symbols are identical (except
from a phase shift of $\pi$ constant over all subchannels) and transmitted
in the coherent mode only for synchronization and channel estimation
purposes.

\section{Implementation Challenges \label{sec:System-Model-1-1}}

In this section, we discuss some real-time implementation issues such
as packet detection and frequency/time synchronization. For each issue,
we also discuss the approaches we followed to address them in the
testbed development.

\subsection{Packet Detection}

% (fold)
\label{sub:packet_detection}

We implement a packet detection scheme that exploits the preamble
repetitions to perform delayed correlation and/or cross correlation
in the time-domain. The delayed correlation operation performs correlation
between the received signal and a delayed copy of itself. However,
the cross correlation operation performs correlation between the known
original preamble and the received signal. The delayed correlation
operation has the advantage of low implementation complexity since
it can be implemented easily via recursive calculations. Moreover,
the performance of the delayed correlation is not degraded in the
presence of the carrier frequency offset (CFO) since it is based on
correlating the received signal with itself. With no CFO, the performance
of the cross correlation scheme is better than the delayed correlation
scheme at the cost of a higher implementation complexity. However,
the performance of the cross correlation scheme is significantly degraded
under CFO.

To achieve a desirable trade-off between complexity and performance,
we employ a hybrid preamble detection technique that exploits both
delayed and cross correlations operations. Specifically, the delayed
correlation scheme can be used as an initial detection stage to obtain
coarse estimates for both packet start and fractional CFO. Note that
CFO estimation/compensation is only needed in the wireless system
since PLC transmissions are in the baseband. In the second stage,
the packet start coarse estimate is refined using a cross correlation
detection process which includes an integer CFO estimation step.

\subsection{Channel Estimation}

In our testbed, to estimate the channel, simple least square (LS)
estimation is implemented in the frequency domain. In particular,
the LS channel estimates are calculated and averaged over all OFDM
symbols in the preamble (nine OFDM symbols) to enhance the LS estimates.
Channel estimation is done once at the beginning of each frame and
used for equalizing the same frame.

\begin{comment}
% (fold)
\label{sub:channel_and_noise_estimation} As with other common OFDM
communication systems, the testbed employs a frequency domain least
square (LS) channel estimator. There are nine OFDM symbols in the
preamble, and the channel estimates obtained from those symbols are
averaged to reduce the noise effects. Channel estimation takes place
once at the beginning of every frame, and the estimate is used to
equalize data symbols in the same frame.
\end{comment}

\subsection{Carrier and Sampling Frequency Offsets}

In our developed testbed, to compensate for carrier and sampling frequency
offsets, both the transmitter and receiver oscillators are synchronized.
In particular, to ensure a negligible carrier frequency offset, a
single local oscillator in the adapter module is used for both up
and down conversion operations. For the sampling frequency synchronization,
a reference clock is generated inside a backplane of a PXI chassis
while the PLC system synchronizes the transceiver sample clocks to
that reference clock. In addition, the transmitter side sends a start
trigger to the receiver side to ensure synchronization between them.
In the wireless system transmission, similar synchronization techniques
are also implemented to ensure perfect synchronization. In our future
work, oscillator synchronization may be disabled while additional
signal processing algorithms will be implemented to eliminate carrier
and sampling frequency offsets.

\begin{comment}
% (fold)
\label{sub:sampling_and_carrier_frequency_offsets} In our testbed
setup, sampling and carrier frequency offsets are compensated by synchronizing
the transmitter and receiver oscillators. In particular, the PLC system
synchronizes sample clocks of the transceiver to a reference clock
generated in a backplane of a PXI chassis to prevent two sample clocks
from drifting from each other. The transmitter also sends out start
triggers to the receiver to ensure that the transmitter and receiver
are synchronized. The wireless system employs an identical synchronization
technique. In addition, a single local oscillator located in the adapter
module is used for both up/down conversion, which results in a zero
(or negligible) carrier frequency offset. These synchronization techniques
eliminate the need for additional signal processing to compensate
for impairments. In our future work, oscillator synchronization will
be removed and compensation algorithms will be employed.
\end{comment}

\subsection{Frame Transmission Synchronization}

In the combining thread, to avoid FIFO queues overflow, the PLC and
wireless systems start frame transmissions at the same time with the
same data rate. Since both the wireless and PLC systems operate independently
(without sharing any triggers/signals), the testbed exploits the deterministic
RTOS advantage and implements a timed-loop in both the wireless and
PLC threads. In particular, the start, idle, and stop flags are generated
in the wireless system then the PLC system reads them through TCP/IP.
Although this may cause a delay (up to several hundred milliseconds),
once both systems initialize, they run at similar speeds with a negligible
time jitter.

\begin{comment}
In order to prevent overflow of the FIFO queues in the combining thread,
frame transmission of the two systems starts at the same time and
transmission rates are the same. However, they operate independently
without sharing any triggers or signals over a wire. The testbed,
thus, takes advantage of the deterministic RTOS and uses a timed-loop
in both the PLC and the wireless threads. The start, idle, and stop
flags are generated from the wireless system and are read by the PLC
system over TCP/IP. This may introduce a delay up to several hundred
milliseconds; however, once the systems start, they run at the same
speed with a negligible jitter
\end{comment}

\section{Conclusion\label{sec:Conclusion}}

We provided an overview of hybrid NB-PLC/unlicensed wireless systems
and discussed different techniques for diversity combining. In particular,
the ASC scheme has the lowest design complexity but worst performance
compared to the PSDC and ISC schemes. On the other hand, the ISC scheme
achieves the best performance but requires much higher design complexity
and pilot overhead. For the best performance/complexity trade-off,
the PSDC scheme is shown to outperform the ASC with a lower design
complexity than the ISC scheme.

In case of known CSI at the transmitter, the TRSD is an appealing
solution. Specifically, TRSD enjoys lower design complexity compared
to the dominant Eigen mode transmission that is typically used for
wireless/wireless or PLC/PLC systems since the latter requires computing
the SVD of the channel matrix in addition to vector multiplications
operations required at the transmit and receive sides. In addition,
unlike wireless/wireless and PLC/PLC diversity techniques, TRSD is
not sensitive to channel or noise estimation errors. Moreover, TRSD
does not require full knowledge about CSI and noise PSD at the transmitter
side (only one bit feedback is required for each OFDM subchannel).
However, for wireless/wireless and PLC/PLC diversity systems, full
knowledge about CSI and noise PSD (per subchannel) is required at
the transmitter.

For differential modulation transmission, the DSSC technique is shown
to be independent of the channel (only the received signal strength
is required) and yet outperforms both the ASC and EGC techniques. 

Finally, our development efforts of a real-time testbed for the evaluation
of PLC/wireless diversity combining techniques were described. The
testbed implements both PLC and wireless systems over CENELEC-A and
unlicensed sub-1 GHz bands, respectively. The transceiver design of
the two heterogeneous systems was implemented to be flexible enough
so that it can adapt to the needs of future research and investigation
in this field.

{\small{}\bibliographystyle{IEEEtran}
\bibliography{SG}
}{\small \par}
\end{document}